\documentclass{aa}  
\usepackage{graphicx}
\begin{document}
   \title{Periodicities in sunspot activity during solar cycle 23}


   \author{Bhuwan Joshi\inst{1}, P. Pant\inst{1} \and P. K. Manoharan
          \inst{2}
          }

   \offprints{B. Joshi}

   \institute{Aryabhatta Research Institute of Observational Sciences (ARIES), 
              Nainital 263 129, Uttaranchal, India\\
              \email{bhuwan,ppant@aries.ernet.in}
         \and
            Radio Astronomy Centre, TIFR, Ooty 643 001, Tamilnadu, India\\
            \email{mano@ncra.tifr.res.in}
              }

   \date{Received ; accepted }

 
  \abstract
  {}   
   {The data of sunspot numbers, sunspot areas and solar flare index during cycle 23 are analyzed
   to investigate the intermediate-term periodicities.}
   {Power spectral analysis has been performed  separately for the data of the 
   whole disk, northern and southern hemispheres of the Sun.}
   {Several significant midrange periodicities ($\sim$175, 133, 113, 104, 84, 63 days) are detected
    in sunspot activity. Most of the periodicities in sunspot numbers generally 
    agree with those of sunspot areas during the solar cycle 23. The study reveals that the periodic variations in the  
    northern and southern hemispheres of the Sun show a kind of asymmetrical behavior.
    Periodicities of $\sim$175 days and $\sim$133 days are highly significant in the sunspot data of 
    northern hemisphere showing consistency with the findings of Lean (1990) during solar cycles 12-21. 
    On the other hand, southern hemisphere shows a strong periodicity of about 85 days in terms of sunspot activity.
    The analysis of solar flare index data of the same time interval does not show any significant
    peak. The different periodic behavior of sunspot and flare activity can be 
    understood in the light of hypothesis proposed by Ballester et al. (2002), 
    which suggests that during cycle 23, the periodic emergence of magnetic flux partly takes place 
    away from developed sunspot groups and hence may not necessarily increase
    the magnetic complexity of sunspot groups that leads to the generation of flares.}
  {}

   \maketitle
%

\section{Introduction}
Sunspots are the fundamental indicators of solar activity. They exhibit a long term periodicity
of 11 years, the so called solar cycle, which is known since a long time. In short term variations,
the 27-day periodicity is the most prominent, which is attributed to the rotation of the Sun. 
In solar cycle 21, a periodicity of 154 days was discovered by Rieger et al. (1984) in the
occurrence of high energy flares. Since then a number of studies have been made to search for the
existence of intermediate-term or midrange periodicities between 27 days and 11 years in the various
features of the active Sun (Bai, 2003).

The presence of near 155 days periodicity in flare and flare related data during cycle 21 was established 
by various authors (Ichimoto et al. 1985; Bogart \& Bai 1985; Bai \& Sturrock 1987;
\"{O}zg\"{u}c and Ata\c{c} 1989; Bai \& Cliver 1990; Bai 2003; Joshi \& Joshi 2005). 
Midrange periodicities in flare data were also investigated 
in solar cycles 19 and 20. Bai (1987) found a 51 days periodicity during cycle 19 in the occurrence
rate of major flares. Analysis of solar flare data of cycle 20 revealed the periodicities of 78 days (Bogart \&
Bai 1985), 84 days (Bai \& Sturrock 1991) and 127 days (Bai \& Sturrock 1991; Kile \& Cliver 1991).
Significant peaks at 74, 77 and 83 days 
were reported in the flare data of solar cycle 22 (Bai 1992; \"{O}zg\"{u}c and Ata\c{c} 1994; Joshi \& Joshi 2005).

In several studies the sunspot data (numbers as well as areas) have been analyzed to investigate the midrange 
periodicities during different solar cycles (Lean \& Brueckner 1989; Lean 1990; Carbonell \& Ballester 1992; 
Oliver et al. 1992; Ballester et al. 1999; Krivova \& Solanki 2002; Richardson \& Cane 2005). 
Lean \& Brueckner (1989) detected the near 155 days periodicity in sunspot blocking function, 
the 10.7 cm radio flux and sunspot number during cycles 19, 20 and 21 but this period could not be
found in plage index. 
Lean (1990) analyzed the data of sunspot areas during cycles 12-21 and 
found that the periodicities in the range of 130 to 185 days occurred intermittently for the interval of 1 to 3 years 
during the epochs of maximum activity. 
Oliver et al. (1998) 
showed a time-frequency coincidence between the occurrence of the periodicity in both sunspot areas and
high energy flares during cycle 21 which suggests a causal relationship between the two phenomena. 

The aim of this paper is to detect midrange periodicities in the data of sunspot numbers, areas and solar 
flare index during solar cycle 23 and to investigate the causal relationship in the periodic behavior of 
these solar activity phenomena. The periodic variations have also been examined separately for 
the northern and southern hemispheres of the Sun.

\section{Data and Analysis}
\subsection{Data}
The daily values of sunspot numbers in the northern and southern hemisphere of the Sun are made available
by National Geophysical Data Centre 
(NGDC) since 1992.
In solar cycle 23, the level of sunspot activity 
was very low during 1996 while in 1997 and 1998 the activity was in the ascending phase (See Joshi \& Pant 2005). 
After January 9, 1998
the daily sunspot numbers never came down to zero value till January 27, 2004, when again no sunspots were
found on the Sun. For the present analysis we have taken the daily data of solar indices from February 1998 to 
December 2003, when there were sunspots visible on each day on the solar disk. Thus our data covers a period
during solar cycle 23 when a moderate to high level of sunspot activity was reported.
The daily data of sunspot areas (expressed in units of millionths of solar hemisphere), 
for both the solar hemispheres (North and South), used in this study are
compiled by Greenwich Observatory/ USNF/ NOAA.
The flare index, introduced by Kleczek (1952), is an excellent parameter to describe the daily flare activity 
on the Sun. Flare index data, analyzed here, are computed by
Kandilli Observatory.

\subsection{Periodogram analysis, normalization and false alarm probability}

   \begin{figure}
   \centering
\includegraphics[width=7.0cm, height=7.0cm]{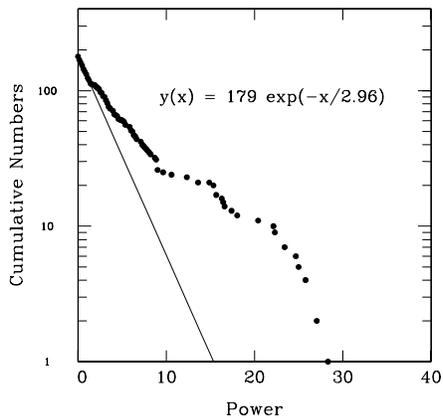}
      \caption{ Scargle power distribution corresponding to the power spectra of daily values of sunspot numbers
for the whole solar disk shown in Figure 2 (top panel). The vertical
axis is the number of frequencies for which power exceeds x. The 
straight line is the fit to the points for lower values of power.
              }
         \label{FigVibStab}
   \end{figure}

We have Used the Lomb-Scargle periodogram method (Lomb, 1976; Scargle, 
1982) modified by Horne and Baliunas (1986) to compute the power spectra
of all the data sets. For a time series X($t_{i}$), the periodogram as a function of frequency $\omega$, 
$P_{N}$($\omega$), is defined in such a manner that if X($t_{j}$) is pure noise then the power in $P_{N}$($\omega$)
follows an exponential probability distribution (Horne and Baliunas, 1986). 
However when each element of time series are not statistically
independent but correlated, the power distribution follows an equation of the form
\begin{equation}
Pr [P_{N}(\omega_{0}) > z] = e^{-z/k},
\end{equation}
where the normalization factor $k$, which is due to event correlation,
should be determined empirically (Bai and Cliver, 1990).
   
   \begin{figure}
   \centering
\includegraphics[width=6.8cm]{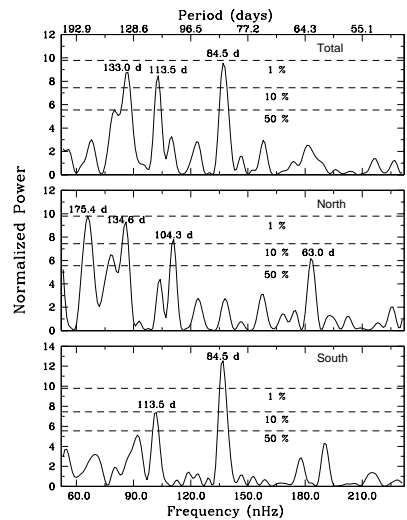}
      \caption{Normalized periodograms of the daily sunspot numbers for the whole disk, 
northern and southern hemisphere of the Sun (from top to bottom) for the time interval from February 1998 to 
December 2003. 
Periodogram is computed
for the frequency interval of 231.5$-$52.6 nHz (50$-$220 days). 
Horizontal dashed lines in 
each panels indicate FAP levels of 1 $\%$, 10 $\%$, and 50 $\%$.}
         \label{FigVibStab}
   \end{figure}

\begin{table*}
\caption{Periodicities detected in sunspot numbers and areas in the total solar disk (T), 
	 northern (N) and southern (S) hemispheres of the Sun during solar cycle 23. Corresponding values of 
	 false alarm probability (\%) are given in brackets.}             
\label{table:1}      
\centering          
\begin{tabular}{l l l | l l l}      
\hline\hline       
\multicolumn{3}{c}{Sunspot Numbers} & \multicolumn{3}{c}{Sunspot Areas} \\ 
\hline                    
T  & N & S & T & N & S \\
\hline
...          & 63 (31.4)  & ...         & ...         & ...         & ...\\
84.5 (1.2)   & ...        & 84.5 (0.06) & ...         & ...         & 85.1 (31.1) \\
...          & 104.3 (7)  & ...         & ...         & ...         & ... \\
113.5 (3.7)  & ...        & 113.5 (10.5)& 112.4 (11.9)& ...         & ... \\
133.0 (2.8)  & 134.6 (1.6)& ...         & ...         & ...         & ...\\
...          & 175.4 (0.9)& ...         & ...         & 175.4 (0.7) & ...\\

\hline
\end{tabular}
\end{table*}

The power spectra for all the data sets is computed for frequency range of 231.5$-$52.6 nHz 
(period interval of 50$-$220 days) with a frequency interval of 1.0 nHz. Therefore 
in each power spectra there are 179 frequencies. In this manner, for example,
the raw periodogram obtained for daily sunspot numbers of whole solar disk showed the highest peak
at 137 nHz (period of 84.5 days) with height 28.03. Next we studied the distribution of power values
corresponding to this raw periodogram which is shown in Fig. 1. 
From Fig. 1, we find that for all 179 frequencies 
the power exceeds zero, we have a point at (X=0, Y=179). At only one frequency 
(137.0 nHz) the power was 28.03, its maximum value. For  lower values of power, the distribution
can be well fitted by the equation y = 179 exp(-x/2.96), as expected
from equation (1). Thus, we normalize the power spectrum by 
dividing the powers by 2.96 to obtain the normalized periodogram which is shown in Fig. 2 (top panel). 
For other cases we use the same procedure for normalization.

The statistical significance of a peak in the power spectrum
is estimated by computing the `False alarm probability' (FAP). 
It is given by the expression
\begin{equation}
F = 1 - [1 - exp(-Z_{m})]^{N},
\end{equation}
where $Z_{m}$ is the height of the peak in the normalized power
spectrum and $N$ is the number of independent frequencies (Horne \& Baliunus 1986; Joshi \& Joshi 2005).
Fourier components calculated
at frequencies at intervals of independent fourier spacing, $\Delta$$f_{ifs}$=$\tau$$^{-1}$, where
$\tau$ is the time span of the data, are totally independent (Scargle, 1982). 
Here we have $\tau$= 2160 days and $\Delta$$f_{ifs}$= 2.14 nHz. 
Thus there are 84 independent frequencies in 231.5$-$52.6 nHz interval. We have oversampled
to obtain this power spectrum in which the normalized height of the peak at 84.5 days is 9.56. The 
oversampling tends to estimate more accurately the peak value.    
Therefore, if we substitute $Z_{m}$=9.56
and N=179 in equation (2) we get $F$=0.0125, i.e., the probability
to obtain such a high peak at 137.0 nHz (84.5 days) by chance is about 1.2$\%$.
The same analysis has been applied to estimate the statistical significance of different peaks 
present in all the power spectra shown in Figures 2, 3 and 4.

To be sure that peaks in the different power spectra (Figs. 2, 3, and 4)
are not due to aliasing, we have removed sine curves of their periods
from the original time series. We find that these peaks get removed
in the power spectra of time series obtained after subtraction.
   
   \begin{figure}
   \centering
\includegraphics[width=6.8cm]{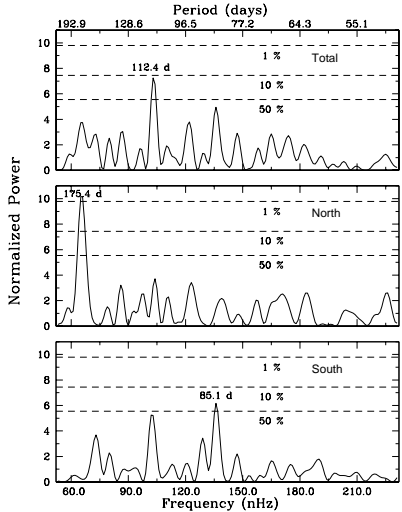}
      \caption{Same as Figure 2, but for daily sunspot areas.}
   \end{figure}

  \begin{figure}
   \centering
\includegraphics[width=6.8cm]{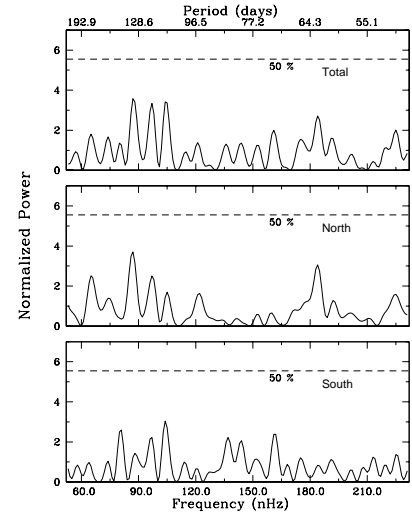}
      \caption{Same as Figure 2, but for daily solar flare index.}
         \label{FigVibStab}
   \end{figure}

\section{Discussion and conclusions}
In this paper, the daily data of sunspot numbers, areas and solar flare index are analyzed to 
search for midrange periodicities.
The periodogram analysis of daily sunspot numbers and areas (Figures 2 and 3) reveals the existence
of several significant periodicities. 
On the other hand, power spectra of daily solar flare index of the same interval (Figure 4)
did not show any prominent peak with FAP value below 50 \%. In table 1, the significant periodicities 
detected in sunspot numbers and areas are summarized.

The sunspot numbers of the whole disk (top panel of Figure 2) shows three important peaks at 84.5 days, 133.0 days
and 113.5 days. A periodicity of 134.5 days is appearing in the sunspot numbers of northern hemisphere 
(Fig. 2 middle panel), which is very close
to 133.0 days period detected in the data of whole disk. However, the northern hemisphere's sunspot numbers exhibit
the highest peak at 175.4 days. Two other important periodicities at 104.3 days and 63 days are found
in northern hemisphere. The sunspot numbers of the southern part of the solar disk gave rise to a highly 
significant peak at 84.5 days (see bottom panel of Figure 2), exactly same peak is detected in whole disk data. Similarly
another peak in south at 113.5 days is also appearing in the whole disk data.

The power spectra of sunspot areas of whole disk, northern and southern hemisphere of the Sun are shown in Figure 3
(from top to bottom panel). In the whole disk data a significant peak at 112.4 days is detected which is very close 
to 113.5 days period present in sunspot numbers of the whole disk as well as southern hemisphere. In can be noticed
that the same peak is also appearing in sunspot areas of southern hemisphere, but its significant level is low.
Similarly 85.1 days periodicity present in southern hemisphere sunspot areas is very near to 84.5 days peak in the 
power spectra of southern hemisphere and full disk values of sunspot numbers. Both the periodicities ($\sim$85 and 113 days)
are absent is northern hemisphere data of sunspot numbers and areas. A periodicity of 175.5 days found
in sunspot areas of northern hemisphere is exactly appearing in sunspot numbers of the same hemisphere and it absent
in total disk as well as southern hemisphere data of both the indices. We can conclude that the near 113 and 85 days 
periodicities in both the indicators of sunspot activity are basically due to the activity in southern part of the solar 
disk, while near 175 days periodicity comes as a result of sunspot activity in northern hemisphere of the Sun.
Here it is interesting to mention that a periodicity of 86 days was detected in sunspot areas of cycle 22
(Oliver \& Ballester, 1995), which is close to 85 days periodicity appeared in  cycle 23.

Highly significant peaks at $\sim$175 and $\sim$133 days detected in the present analysis in sunspot 
numbers are consistent with the studies of Lean (1990) and Richardson \& Cane (2005). The Periodogram analysis 
of sunspot areas further confirms the presence of near 175 days periodicity. Lean (1990) made an
extensive study of periodic patterns in sunspot areas during cycles 12-21, and found that the 
``Rieger-type" periodicities occur intermittently in each cycle during the epochs of maximum activity, however
its actual period varies from 130 to 185 days. 
Richardson \& Cane (2005) applied wavelet analysis to 
sunspot number data of cycle 23 and reported the appearance of periodic signals of enhanced power  
in the range of $\sim$120-200 days during the maximum phase of the cycle (late 1998 to 2002). 

The cause for the origin of Rieger-type periodicities is suggested by several authors. Lean \& Brueckner (1989)
linked this periodicity with the magnetism of sunspots and suggested that it may be a property of emerging 
flux rather than the total amount of flux present on the solar disk. Carbonell \& Ballester (1990) 
suggested its association with the periodicity in emergence of magnetic flux through the photosphere.
Ballester et al. (2002, 2004) analyzed the photospheric magnetic flux during cycles 21-23 and 
found the evidence of near 160 days periodicity in cycles 21 and 23. Based on their analysis, they 
put forward a hypothesis that the periodicity can appear in two different ways. 
Periodic emergence of magnetic flux within already formed sunspot groups enhances the magnetic complexity of the active
region which gives rise to similar periodic variations in solar flares. 
On the other hand, the periodic emergence of flux away from developed sunspot groups will not be reflected
in the periodic behavior of solar flares.

The comparison of periodogram analysis of three kinds of data sets during February 1998 to 
December 2003, presented in this paper,
suggests that the results of sunspot numbers
generally agree with the results of sunspot areas. However the periodic pattern of occurrence of solar 
flares displayed a different
behavior and we could not detected any significant periodicity in solar flare index data of same time interval.  
Recently Ata\c{c} \& \"{O}zg\"{u}\c{c} (2006) analyzed the solar flare index data  
from January 1999 to December 2002 and detected a $\sim$126 days periodicity. 
Their analysis also did not show any periodicity in the range of 130 to 185 days which was found in 
our study of sunspot data. Similar results were obtained with the 
analysis of soft X-ray flare index (Joshi \& Joshi 2005). 
It therefore seems that periodic emergence of magnetic flux away from developed active region, during
the maximum phase of current solar cycle, caused the similar periodic variations in sunspot numbers as well as
areas, but did not enhance the magnetic complexity of the active regions in the same proportion.
This may be a reason why the similar periodic variations could not be detected in solar flare index data.

\begin{acknowledgements}
We thank Prof. Ram Sagar for useful discussions and encouragement. Flare Index data used in this study were calculated by Dr. T. Ata\c{c} and Dr. A. \"{O}zg\"{u}\c{c} from Bogazici University Kandilli Observatory, Istanbul, Turkey. The authors owe
sincere thanks to the referee, Dr. I. G. Richardson, for very useful comments and suggestions 
which helped to improve the paper
significantly. Useful discussion with Dr. R. Oliver, Universitat de les Illes Balears, Spain about the 
periodogram analysis is also gratefully acknowledged.

\end{acknowledgements}


\begin{thebibliography}{}
   \bibitem[2006]{atac} Ata\c{c} \& \"{O}zg\"{u}\c{c} 2006,
      Sol. Phys., 233, 139

   \bibitem[1987]{bai} Bai, T. 1987, 
      ApJ, 318, L85
   
   \bibitem[1987]{bai} Bai, T. \& Sturrock 1987, 
      Nature, 327, 601
   
   \bibitem[1991]{bai} Bai, T. \& Sturrock 1991, 
      Nature, 350, 141
   
   \bibitem[1992]{bai} Bai, T. 1992, 
      ApJ, 388, L69
   
   \bibitem[2003]{bai} Bai, T. 2003,
      ApJ, 591, 406
   
   \bibitem[1990]{bai} Bai, T. and Cliver, E. W. 1990, 
      ApJ, 363, 299
  
   \bibitem[1999]{ballester} Ballester, J. L., Oliver, R., \& Baudin, F. 1999,
      ApJ, 522, L153
   
   \bibitem[2002]{ballester} Ballester, J. L., Oliver, R., \& Carbonell, M. 2002,
      ApJ, 566, 505
   
   \bibitem[2004]{ballester} Ballester, J. L., Oliver, R., \& Carbonell, M. 2004,
      ApJ, 615, L173
   
   \bibitem[1985]{bogart} Bogart, R. S. and Bai, T. 1985, 
      ApJ, 299, L51
   
   \bibitem[1990]{carbonell} Carbonell, M. \& Ballester, J. L. 1990,
      A\&A, 238, 377
   
   \bibitem[1992]{carbonell} Carbonell, M. \& Ballester, J. L. 1992,
      A\&A, 255, 350
   
   \bibitem[1986]{horne} Horne, J. H. and Baliunas, S. L., 1986,
      ApJ, 302, 757

   \bibitem[1985]{ichimoto} Ichimoto, K., Kubota, J., Suzuki, M., Tohmura, I., and Kurokawa, H. 1985, 
      Nature, 316, 422
   
   \bibitem[2005]{joshi} Joshi, B. \& Joshi, A. 2005,
      Sol. Phys., 226, 153
   
   \bibitem[2005]{joshi} Joshi, B. \& Pant. P. 2005,
      A\&A, 431, 359
   
   \bibitem[1991]{kile} Kile, J. N. \& Cliver E. W., 1991,
      ApJ, 370, 442
   
   \bibitem[1952]{Kleczek} Kleczek, J.: 1952, 
    Publ. Contr. Astron, No 22, Prague
   
   \bibitem[2002]{krivova} Krivova, N. A. \& Solanki, S. K. 2002,
      A\&A, 394, 701
  
   \bibitem[1989]{lean} Lean, J. L. \& Brueckner, G. E. 1989,
      ApJ, 337, 568
 
   \bibitem[1990]{lean} Lean, J. L. 1990,
      ApJ, 363, 718
   
   \bibitem[1976]{lomb} Lomb, N. 1976,
      Ap\&SS, 39, 477
   
   \bibitem[1992]{oliver} Oliver, R., Carbonell, M. \& Ballester, J. L.  1992,
      Sol. Phys., 137, 141
   
   \bibitem[1995]{oliver} Oliver, R., \& Ballester, J. L.  1995,
      Sol. Phys., 156, 145
    
   \bibitem[1998]{oliver} Oliver, R., Ballester, J. L., and Baudin, F. 1998,
      Nature, 394, 552
   
   \bibitem[1992]{ozguc} \"{O}zg\"{u}\c{c}, A. and Ata\c{c}, T. 1989,
      Sol. Phys., 123, 357
   
   \bibitem[1994]{ozguc} \"{O}zg\"{u}\c{c}, A. and Ata\c{c}, T. 1994,
      Sol. Phys., 150, 339
   
   \bibitem[1984]{Rieger} Rieger, E., Share, G. H., Forrest, D. J. et al. 1984,
    Nature, 312, 623
   
   \bibitem[2005]{Richardson} Richardson, I. G. \& Cane H. V. 2005,
    \grl, 32, L02104

   \bibitem[1982]{Scargle} Scargle. J. D.: 1982, 
      ApJ, 263, 835

\end{thebibliography}
\end{document}